\documentclass[10pt,twocolumn,epsfig]{nature}
\usepackage{epsfig}
\usepackage{graphicx}

\title{Giant Rashba effect at the topological surface of PrGe revealing antiferromagnetic spintronics} 

\begin{document}

\author{Soma Banik$^{1*}$, Pranab Kumar Das$^{2}$, Azzedine Bendounan$^3$, Ivana Vobornik$^4$, A. Arya$^5$, Nathan Beaulieu$^3$, Jun Fujii$^4$, A. Thamizhavel$^2$, P. U. Sastry$^6$, A. K. Sinha$^1$, D. M. Phase$^7$,  S. K. Deb$^1$}

\maketitle
\begin{affiliations}
\item{Indus Synchrotrons Utilization Division, Raja Ramanna Centre for Advanced Technology, Indore, 4520 013, India.}
\item{Department of Condensed Matter Physics and Materials Science, Tata Institute of Fundamental Research, Homi Bhabha Road, Colaba, Mumbai 400 005, India.}
\item{Synchrotron SOLEIL, L$'$Orme des Merisiers, Saint-Aubin, BP 48, FR-91192 Gif-sur-Yvette Cedex, France.}
\item{Istituto Officina dei Materiali (IOM)�CNR, Laboratorio TASC, Area Science Park, S.S.14, Km 163.5, I-34149 Trieste, Italy.}
\item{Materials Science Division, Bhabha Atomic Research Centre, Mumbai 400085, India.}
\item{Solid State Physics Division, Bhabha Atomic Research Centre, Mumbai 400085, India.}
\item{UGC-DAE Consortium for Scientific Research, Khandwa Road, Indore 452001, India.}
\end{affiliations} 
\begin{abstract} 
 
Rashba spin-orbit splitting in the magnetic materials opens up a new perspective in the field of spintronics. Here, we report a giant Rashba-type spin-orbit effect on PrGe [010] surface in the paramagnetic phase with Rashba coefficient $\alpha_R$=~5~eV$\AA$. Significant changes in the electronic band structure has been observed across the phase transitions from paramagnetic to antiferromagnetic (at 44~K) and from antiferromagnetic to the ferromagnetic ground state (at 41.5~K). We find that Pr $4f$ states in PrGe is strongly hybridized with the Pr $5d$ and Ge $4s-4p$ states near the Fermi level which is responsible for the presence of Rashba effect. The behavior of Rashba effect is found to be different in the $k_x$ and the $k_y$ directions showing electron-like and the hole-like bands, respectively. The possible origin of Rashba effect in the paramagnetic phase is related to the anti-parallel spin polarization present in this system. First-principles density functional calculations of Pr terminated surface with the anti-parallel spins shows a fair agreement with the experimental results. We find that the anti-parallel spins are strongly coupled to the lattice such that the PrGe system behaves like weak ferromagnetic system. Analysis of the energy dispersion curves at different magnetic phases showed that there is a competition between the Dzyaloshinsky-Moriya interaction and the exchange interaction which gives rise to the magnetic ordering in PrGe. Supporting evidences of the presence of Dzyaloshinsky-Moriya interaction are observed as anisotropic magnetoresistance with respect to field direction and first-order type hysteresis in the X-ray diffraction measurements. A giant negative magnetoresistance of 43$\%$ in the antiferromagnetic phase and tunable Rashba parameter with temperature across the magnetic transitions makes this material a suitable candidate for technological application in the antiferromagnetic spintronic devices.    
\end{abstract}

\maketitle
\section*{Introduction}
One of the main challenges in the field of spintronics is to manipulate the spin structures by electric and spin currents. In this context the materials with Rashba spin-orbit (SO) coupling \cite{Rashba} have gained a huge amount of interest due to their use in the Spintronics technology\cite{Fabian04, Fabian07}. Rashba effect in the non-magnetic systems has been extensively studied \cite{Nitta, Popovic, Bendounan, Ast} but there are very few reports on the magnetic metallic systems showing the Rashba effect\cite{Krupin05, Krupin09,Brataas12}. In ferromagnetic (FM) materials with the SO splitting the magnetization reversal gives rise to Rashba effect which can be used for the giant magnetoresistance devices and spin transfer torque device applications.\cite{Krupin09, Brataas12} On the other hand, antiferromagnetic (AFM) materials are the hidden magnets and the magnetism in these systems is difficult to probe experimentally. The magnetization reversal in the AFM systems will not make any change because of zero net magnetic moment. In such a case, the rotation of the spins will lead to the change in the resistance and has important technological applications \cite{Wang14, Fina14}. The anisotropic magnetoresistance in these AFM systems is linked to the SO interaction. Hence, the AFM systems which have significant anisotropy can be the promising candidates for the AFM spintronic device applications \cite{Barthem}.    

In recent years there has been flurry of research in understanding the anisotropic magnetic properties of several Pr-based compounds \cite {Joshi,Janssen, Kitai, Suzuki, Mulders}. Pr-based compounds exhibit a variety of ground state properties due to the critical role of crystalline-electric-field effects, quadrupolar fluctuations ${\it etc}$ \cite{Joshi, Janssen, Kitai, Suzuki, Mulders}. The mechanism of the heavy Fermion behavior in Pr-compounds is quite different from the usual Kondo route to heavy Fermions in Ce-based compounds.  The crystal structural, transport and magnetic properties of PrGe single crystal have been studied in great detail \cite{Das, Hohnke, Buschow}. PrGe crystallizes in CrB type crystal structure with $Cmcm$ space group \cite{Das}. Magnetic studies on PrGe single crystal showed two magnetic transitions at 44~K related to the AFM transition and 41.5~K related to the FM transition \cite{Das}. Interestingly, a higher effective magnetic moment of Pr in PrGe has been observed for [010] crystallographic orientation $\sim$3.90~$\mu_B$, which is not only higher than the effective moment of free Pr$^{3+}$ ion $\sim$3.58~$\mu_B$ but also higher than the moment observed along [100] ($\sim$3.78~$\mu_B$) and [001] ($\sim$3.71~$\mu_B$) crystallographic orientations. Since, the effective magnetic moment is related to the magnetocrystalline anisotropy (MCA), which is intrinsically linked to the SO interaction that couples the spins to the symmetry of the lattice. Hence, to understand the origin of different magnetic ground states in PrGe it is utmost important to investigate the nature of SO interaction present in this system. 

In the present work, we have investigated the band structure and electronic density of states by angle resolved photoemission (ARPES) and resonant photoemission measurements, respectively. We report that the SO splitting observed on the topological surface of PrGe [010] single crystal is due to the fact that the spins of Pr atoms are anti-parallel and perpendicular to the surface. Across the magnetic transition, the strong SO coupling gets disturbed due to the hybridization between the Pr $4f$ electrons, with the Pr $5d$ and Ge $4s-4p$ conduction electrons which leads to the change in the electronic band structure. The canted anti-parallel spin gives the evidence that the Dzyaloshinsky-Moriya (DM) interaction \cite{Heide} is present in this system and the coupling between the electron-like and the hole-like bands is responsible for the origin of different magnetic phases in PrGe.

\section*{Results and Discussions}


High resolution ARPES measurements have been performed to determine the electronic band structure of PrGe [010] single crystal at $h\nu$=~28~eV. Figs. 1 (a) and (b) show the electronic bands in the paramagnetic (PM) phase at 100~K and FM phase at 20~K, respectively. The most interesting observation in the band structure is that a band at 0.12~eV shows a Rashba type SO splitting in the PM phase (Fig. 1 (a)) near the $\Gamma$ point. The spin split bands have hole-like character. The effect of Rashba SO splitting is significantly less in the FM phase (Fig. 1 (b)). No major changes are observed for the bands lying at higher energies ($\geq$-0.5~eV). The changes observed in these bands across the phase transition from PM to FM phase clearly indicates that the surface states get modified across the magnetic transition in this system. In other magnetic systems like Gd, Tb ${\it etc.}$ where in-plane magnetic moment is present, the exchange splitting is reported to be dominant over the Rashba effect. In such a case, the Rashba SO split bands are observed only when the magnetization direction is reversed.\cite{Krupin05, Krupin09} In the present case the Rashba SO split bands in the PM phase are observed without any in-plane magnetization and hence, it must have a different origin which need to be explored.
 
To understand the role of surface states and their correlation with the magnetism present in this system, we have performed a detailed band structure study as function of temperatures and shown in Figs. 1 (c) to (l). The behavior of the split bands in the PM phase remains similar (Figs. 1 (c) to (g)) between 100~K to 60~K. A drastic change is observed in the bands at and below 51~K (Fig. 1(h)) which is close to the AFM transition. In the AFM phase, the spin split feature in the hole-like bands disappears and is found to be overlapped with the electron-like bands. On further lowering the temperature, in the FM phase, we have observed both the overlapped electron-like and hole-like bands with prominent spin split features (Figs. 1 (j), (k) and (l)). The changes in the band structure across the magnetic transition may be due to the mixing of the Pr $4f$ states with the other valence states in this system. Similar kind of changes in the band structure have been observed in CeSb across the PM to AFM phase transition \cite{Kumigashira97} which was attributed to the mixing of Sb $5p$ bands with the Ce $4f$ bands.


We find that the Rashba type SO splitting is large in the PM phase. Hence, to understand the character of the bands which give rise to the splitting, we have performed the resonant photoemission (RPES) measurements at room temperature. RPES has been performed across the Pr $4d-4f$ resonance in the photon energy range from 110 to 140~eV. The RPES data are plotted in the contour plot as shown in Figs. 2 (a), where 4 prominent features marked as A, B, C and D at -0.65, -3.2, -5 and -8 eV respectively, are observed. Across the Pr $4d$ to $4f$ resonance, feature $A$ shows a significant enhancement in intensity (Fig. 2(a)). The constant initial state (CIS) spectra for the feature $A$ at -0.65~eV has been plotted in the inset of Fig. 2(a) using the standard method discussed elsewhere \cite{Soma1, Soma2}. To understand the character of this feature Fano line profile \cite{Fano} of the form $\sigma(h\nu)=\sigma_a \frac{(q+\epsilon)^2}{1+\epsilon^2} + \sigma_b$ and $\epsilon=(h\nu-E_0)/W$ has been fitted and shown in the inset of Fig. 2(a) with a solid line. Here, the parameters $E_0$, $q$ and $W$ represents the resonance energy, discrete/continuum mixing strength and the half-width of the line. The value of the parameters determined from fitting are $E_0$=~125~$\pm$~0.02~eV, $q$=~3.69~$\pm$~0.01 and $W$=~1.92~$\pm$~0.01~eV. The larger value of $q$ indicates that the states at -0.65~eV BE (feature A) is localized in nature. Two broad features B and C have been observed between -0.5 to -4.5~eV which do not show the resonance. We find that the Pr $4d-4f$ resonance in PrGe is quite different from the results reported for thick Pr films \cite{Hwang97}. The localized Pr $4f$ states at -3.6~eV in bulk Pr films show a larger enhancement than the features near $E_F$ which are mainly the Pr $5d$ states\cite{Hwang97}. However, in PrGe the enhancement of the states near E$_F$ gives a clear indication that both the localized and the itinerant character of the Pr $4f$ states play important role in the magnetism of this system. Similar enhancement of Pr $4f$ states near E$_F$ has been observed in the RPES spectra of high $T_C$ superconductor Y$_{1-x}$Pr$_x$Ba$_2$Cu$_3$O$_{7-\delta}$ \cite{Kang89} which has been attributed to the hybridization of the Pr $4f$ states with the Cu $3d$ and O $2p$ states in this system. For PrGe, the resonance has been observed at $\sim$125~eV which is much above the Pr $4d$ threshold energy (114~eV) hence confirms that there is a finite hybrization present in this system. $E_0$ ($\sim$125~eV) of Pr $4f$ states in the PrGe matches well with the reported value of the $E_0$ for thick Pr films \cite{Hwang97} and high $T_C$ superconductor Y$_{1-x}$Pr$_x$Ba$_2$Cu$_3$O$_{7-\delta}$ \cite{Kang89}.

The signature of hybridization between the $4f$ states and the conduction electrons is well known to give rise to features in the core level spectrum and is studied in detail in other rare-earth based systems \cite {Fuggle, Soma1}. Pr $3d$ core level spectrum is shown in Fig. 2(c) where SO splitting of 20.1~eV has been obtained between the Pr $3d_{5/2}$ and $3d_{3/2}$ peaks. Extra features marked as $f^3$ and $m$ in Fig. 2 (c) have been observed for both the Pr $3d_{5/2}$ and $3d_{3/2}$ peaks. The asymmetric feature $m$ arises due to the multiplet electronic states and has also been seen in other Pr based systems \cite{Fujimori84}. The main peak $f^2$ is associated with the poorly screened $3d^9f^2$ states while the other feature marked as $f^3$ arises from the $3d^9f^3$ configuration \cite{Felner92,Chung01}. $f^3$ feature has been attributed to $5d\rightarrow4f$ satellite \cite{Felner92,Chung01} which further confirms the hybridization of the Pr $4f$ states with the conduction electrons.

In Fig. 2(b), the difference of the on-resonance and the off-resonance spectra at 116~eV and 125~eV are shown. The positive part in the difference spectrum (shaded yellow color) shows the positions of the partial density of Pr $4f$ states and negative part shows the contribution from other states present in the valence band (VB). We have performed the first principles density of states (DOS) calculations within DFT using GGA method by considering the experimental lattice parameters\cite{Das} and the bulk structure as shown in Fig. 2. The SO coupling for the Pr $4f$ states have been included in the calculations. The partial DOS (PDOS) of Pr $4f$, Pr $5d$, Ge $4s$ and Ge $4p$ states are shown in Fig. 2(d), (e), (f) and (g), respectively. PDOS of Pr $4f$, Ge $4s$ and Ge $4p$ states show a large intensity at -1.5~eV below $E_F$. At $E_F$, Pr $5d$ states show a larger intensity. A standard method of broadening the DOS as mentioned elsewhere \cite{Soma1,Soma2} has been adopted for comparing the experimental and theoretical PDOS of Pr $4f$ states as shown in Fig. 2(h). We find that the bulk calculation does not match with the experimentally obtained PDOS of Pr $4f$ states (Fig. 2(h)).

To understand the discrepancy between the experimental and the bulk theoretical PDOS of Pr $4f$ states in PrGe it is important to understand the nature of Pr spin states which gives rise to the magnetic ordering in this system. It is known that elemental Pr has a stable paramagnetic phase down to 1K. However, magnetic ordering in PrGe is reported due to the crystal field splitting which results in a doublet state in this system \cite{Das}. Since, Pr $4f$ state has 3 electrons and there are 4 possible degenerate states, which mean one spin that is uncompensated and coupled strongly to the orbit will play a major role in the magnetism of the system. The presence of Rashba type spin orbit splitting in the PM phase in PrGe indicates that there is a strong interaction of the uncompensated spin of one Pr atom with the other Pr atom. We find that PrGe behaves similar to a weak ferromagnetic systems in which magnetism arises mainly because the magnetic atoms in different sublattices have antiparallel spin arrangements. The antiparallel spins which are strongly coupled to the orbit gives to the Rashba effect at the surface in this system. To explore the above mentioned possibilities, we have performed the DOS calculation of [010] surface of PrGe. We have used the periodic supercell or slab model approach \cite{Arya03}. The vacuum layers of thickness of 10~$\AA$ were added to the surface to minimize interatomic interactions between periodic images of the slabs. The slab was periodically repeated in three dimensions to facilitate calculations in reciprocal space. For our calculations, we have selected Pr-terminated PrGe [010] surface (see Fig. 2) in the anti-parallel spin configuration (for Pr atoms) incorporating spin-orbit coupling. All the atoms in the surface slab were fully relaxed. Our relaxed positions of atoms indicated deviation of less than $\pm$~0.001 $\AA$ from the initial positions. We have shown the PDOS of Pr $4f$ and Pr $5d$ states of the top Pr atoms in the surface slab calculation in Fig. 2(d) and (e). Compared to the bulk PDOS, we find that the Pr $4f$ and the Pr $5d$ states has a larger intensity at and near $E_F$ (Fig. 2(d) and (e)) in the surface slab calculation. The energy position of the PDOS of all the states remain same in both the bulk and the surface slab calculation except for the Ge $4p$ state, which show a 0.46~eV shift towards higher energy for the surface calculation as compared to the bulk calculation (Fig. 2(g)). The Ge $4s$ PDOS (Fig. 2(f)) shows a slightly higher intensity in the surface calculation. The broadened PDOS of Pr $4f$ states obtained from the surface slab calculation shows a good agreement with the experimental  PDOS (Fig. 2(h)). We find that the total magnetic moment obtained for the top Pr atom in the surface slab calculation is very small about 0.009 $\mu_B$ which further supports the argument that the PrGe system is indeed similar to weak ferromagnetic systems.  

It is well known that in a weak ferromagnetic system, the magnetism arises due to the competition between the DM interaction and the isotropic ferromagnetic exchange interaction. DM interaction is an antisymmetric exchange interaction originating from the SO coupling and leads Rashba effect observed in this system. Although, many theoretical works reported the DM interaction but still there is lack of experimental evidences\cite{Barnes14} to show its presence in a system. This has motivated us to further understand the energy dispersion curves in presence of DM interaction.


In Fig. 3(a), (b) and (c), we show the band structure in the PM (100~K), near AFM (50~K) and FM (20~K) phases respectively in $k_y$ direction and in Fig. 3 (d) and (e) the band structure are shown along the $k_x$ direction in the FM and the PM phase at 20~K and 100~K, respectively. It is to be noted that the Rashba spin-split bands in $k_x$ direction (Fig. 3 (d) and (e)) are electron-like bands. In presence of DM interaction\cite{Barnes14} the single particle energy for the magnetic system is: $E_{k\sigma}~=~\frac{\hbar^2}{2m^*}[(k_x - \sigma K_0 sin\theta)^2 + k_y^2]- E_R sin^2\theta - \sigma[(J_0S)^2 + \alpha_R^2(k_x^2cos^2\theta + k_y^2)]^{1/2}$. Here,$\sigma$ is the carrier spin index=~$\pm$1, $\theta$ denotes the angle of the magnetization with respect to the $Z$-direction, $m^*$ is the effective mass and $J_0S=~\Delta$ represents the coupling between the magnetic impurity and the carriers where $J_0$ is the exchange strength and $S$ is the spin of the magnetic impurity.$\alpha_R$ denotes the coupling constant in the SO Hamiltonian which is described by $\frac{\hbar^2 k_0}{m^*}$. Rashba energy $E_R$, is described by $\frac{m^*\alpha_R^2}{2\hbar^2}$. We have stimulated the energy dispersion curves considering three major conditions for DM interaction in this system: 1) magnetization axis ($M$) is exactly perpendicular to the surface, in this case $\theta$=0, 2) $M$ is along $Z$-axis but tilted in $XY$-plane and 3) $M$ is parallel to the $Y$-axis and the direction of moment tilt away from the $Y$-axis in the direction perpendicular to surface, in this case $\theta$=90$^\circ$. The detail discussion on the stimulations are shown in the supplementary information. 

In the experimental band structure (Fig. 1 (c) to (l)) it is quite clear that the shift of the bands in the momentum space from the $\Gamma$ point is almost constant however there is a drastic changes in the shape of the bands across the magnetic phase transition. Hence, in the stimulation as shown by the solid lines in Fig. 3(a), (b) and (c) we have varied the value of $k_0$, $m^*$, $\Delta$ and $\theta$¸ to understand their actual effect on the band structure. We find that the experimental observations matches very well with the third condition. The reason for observing hole-like bands in the $k_y$ direction and electron-like bands in the $k_x$ direction is because $M$ is parallel to $Y$-axis (supplementary information Fig. 2(c)). In such a case, the exchange field is along the $Y$-axis gets coupled with the kinetic energy of the electrons which causes the Fermi sea to shift along the $X$-axis\cite{Barnes14} (supplementary information Fig. 3 (g), (h), (i)). From the analysis performed on the experimental band structure, we obtain the value of $k_0$ $\approx$~0.11~$\pm$~0.005~$\AA^{-1}$ along the $X$-axis which is same in both the PM and FM phase. The value of $k_0$ in PrGe [010] is about 7.3 times higher than the value of $k_0$$\approx$~0.015~$\AA^{-1}$ as reported for Gd [0001]\cite{Krupin05, Krupin09}. The Fermi sea in the $k_x$ direction is lowered by an energy $E_R$$\approx$~0.58~$\pm$~0.002~eV ${\it w.r.t}$ $E_F$. The value of $m^*$ is negative for the hole-like bands and positive for the electron-like bands. The important result obtained along the $k_y$ direction for the hole like bands is that the effective mass $m^*$ is found to increase in the FM phase (0.466~$\pm$~0.005~$m_e$)and in the AFM phase (0.326~$\pm$~0.005~$m_e$) than in the PM phase (0.176~$\pm$~0.005~$m_e$). In the $k_x$ direction the effective mass of the electron-like states is found to be two times less as compared to the hole-like states observed in the $k_y$ direction for that particular phase. The value of $\alpha_R$ obtained is $\approx$~5~$\pm$~0.02~eV$\AA$ at 100~K. This is probably the largest value of $\alpha_R$ obtained so far for the metallic surface states. We find that the value of $\alpha_R$ decreases across the phase transitions from $\sim$2.7~$\pm$~0.02~eV$\AA$ in near AFM phase to $\sim$1.89~$\pm$~0.02~eV$\AA$ in the FM phase. The highest value of $\alpha_R$ in the paramagnetic phase over a wide temperature range from 300~K to 60~K makes it an important material for spintronics applications. We also report that the Rashba parameter $\alpha_R$ in PrGe can be tuned by varying the temperature. $\Delta$ mainly causes a shift in the spin-up and the spin-down bands in the near AFM phase and in the FM phase which is quite evident in Fig. 3(b) and (c). Highest value of $\Delta$ $\approx$0.120~eV is obtained for the AFM phase (50~K, Fig. 3(b)) which reduces to $\approx$0.030~eV in the FM phase (20~K Fig. 3(c)). Spin canting with respect to the $Y$-axis (Fig. 3(g, supplementary information Fig. 2 (c)) can be determined from the value of $\alpha_R$ and $\Delta$ such that the canting angle $\delta(k_y)$=~$tan^{-1}$($\alpha_R$$(k_y)$/$\Delta$). The value of canting angle is almost $90^\circ~\pm~0.02^\circ$ in the PM phase, $87.5^\circ~\pm~0.02^\circ$ in the near AFM phase and $89.1^\circ~\pm~0.02^\circ$ in the FM phase. The appearance of electron like states in the near AFM and the FM phase is due to the finite interaction between the hole-like and the electron-like states which not only increase the effective mass of the electrons across the phase transition but also results in the magnetic ordering in this system. 

To further understand the behavior of the bands in both the $k_y$ and $k_x$ direction, we have performed the theoretical band structure for the surface slab calculation as discussed earlier. The total magnetic moment obtained for the top Pr atom in the surface slab calculation is very small about 0.009 $\mu_B$. The orientation of the moment are considered to be perpendicular to [010] surface of PrGe (Fig. 3(f)) which is similar to the condition where the moment is tilted maximum ($\delta$=$90^\circ$) with respect to the $Y$-axis. The Pr terminated surface of the PrGe which has spin orientation perpendicular to the surface is shown in Fig. 3(f) and the brillouin zone is shown in Fig. 3(g). The 0~k band structure calculation for the Pr terminated surface along the $X'$-$\Gamma$-$X$ and $Z'$-$\Gamma$-$Z$ direction are shown in Fig. 3 (h) and Fig. 3 (i). On comparing the theoretical and experimental band structure at 20~K along the $k_y$ direction (Fig. 3(h) and Fig. 3(c)) and along the $k_x$ direction (Fig. 3(i) and Fig. 3(d)) a fair agreement has been obtained. The theoretical band structure also show interacting electron-like and hole-like bands near the $\Gamma$ point in the $k_y$ direction and the electron-like bands in the $k_x$ direction as observed in the experimental band structure. Small differences in the experimental and theoretical band structure has been related to the orientation of the spins which is actually tilted in the FM phase ($\delta$=89.1$^\circ$)in the experiment. 


For further understanding the spin structures present in this system, we have performed the magnetoresistance (MR) measurements. MR as a function of temperature is shown in Fig. 4 (a) and (b). Negative MR has been observed in the PM phase at 60~K (Fig. 4(a)), which clearly indicates that there is a net magnetization present in this system. A large MR of 43$\%$ at 8~T field is observed in the AFM phase at 42~K. However in the FM phase 30~K the MR is more positive than the AFM phase. The reason for obtaining a large MR in the AFM phase than in FM phase could be related to the exchange splitting and the canted spin orientation which is more enhanced in the AFM phase than in the FM phase (Fig. 3(b) and (c)). Similar behavior in the magnetoresistance has been observed for PrGa, where a large magnetoresistance of 34$\%$ is reported at 5~T field in AFM phase.\cite{Chen11} Also it has been observed that the MR decreases and becomes more positive in the FM phase in PrGa which is attributed to the change in the lattice parameters during AFM to FM phase transition on the application of field.\cite{Chen11} Another interesting result in PrGe is that an anisotropic MR behavior ${\it w.r.t.}$ field direction (Fig. 4(b)) is observed at all the temperatures only for the [010] crystallographic orientation (supplementary information Fig.4). Similar anisotropic behavior in MR has been observed in the Pt/Co/Pt films \cite{Franken14}, CrO2 thin films\cite{Anwar13} and MnSi nanowires \cite{Du15} where it is related to the chiral magnetic structure present in these systems. The chiral magnetic structure lacks inversion symmetry and has a strong spin-orbit coupling which is mainly induced by the DM interaction. It is well known that the origin of the chiral magnetic states are related to the change of lattice structure at the crystal boundary or due to the surface contribution which generates an additional magnetic anisotropy.       

DM interaction is expected to influence the lattice structure, hence we have performed temperature dependent x-ray diffraction (XRD) studies on the powder sample to probe its existence. The XRD data have been recorded at few temperatures using laboratory based (Cu K$_\alpha$) source (Fig. 4 (c))  and linearly polarized synchrotron source (13~KeV energy) (Fig. 4(d)). XRD patterns from lab source did not show any change in the peak profiles till the lowest temperature. This indicates no structural change across the magnetic transition and the sample retains CrB type structure with orthorhombic space group Cmcm (inset of Fig. 4(c)). 

XRD patterns from the polarized synchrotron source showed some additional peaks as compared to those recorded from lab source. In Fig. 4(d) the strong peak (around 18.7$^\circ$) shows signatures of splitting into two peaks at 18.64$^\circ$ and 18.71$^\circ$ (shown by dotted lines in the Fig. 4(d)). Similar behavior in the split peaks has been observed in both heating and cooling cycles. On plotting the intensity of the 18.97$^\circ$ peak as a function of temperature (Fig. 4(e)), we find a hysteresis around the magnetic transition. Similar kind of hysteresis as a function of temperature has been observed for other peaks, which indicates the first order nature of magnetic transition that is in good agreement with our earlier work on PrGe \cite{Das}. DM interaction induces the first order magnetic phases is reported in MnSi \cite{Janoschek13}.Hence, the magnetic transition in this system influence the lattice structure due to strong coupling between the spins and the lattice which further confirms the presence of DM interaction in this system.

So, following conclusions are drawn from the present work: 1) PrGe behaves like a weak ferromagnetic system in which the Pr atoms in different sublattices has antiparallel spin orientation and the evidence of strong coupling between the spins and the lattice has been obtained in experiments. 2) Origin of giant Rashba effect on the PrGe [010] surface in the PM phase is due to the breaking of space inversion symmetry by the DM interaction. 3) DM interaction arises in this system due to the strong SO coupling of the magnetic Pr $4f$ electrons and its hybridization with Pr $5d$ and Ge $4s-4p$ conduction electrons. 4) Magnetic ordering both ferromagnetic and antiferromagnetic arises due to the competition between the DM interaction and the exchange interaction present in this system which causes the canting of spins. 5) In the FM and AFM phases, we find that there is a finite interaction between the hole-like and the electron-like states which enhances the effective mass of the electrons at these phases as compared to the PM phase. Effective mass of the electrons has a large influence on the Rashba parameter and found to decrease in the FM and AFM phase as compared to the PM phase. Hence, at the PrGe [010] surface we not only obtain giant Rashba effect ($\alpha_R$=~5~eV$\AA$) in the PM phase, giant negative MR (43$\%$) in the AFM phase but also there is possibility to tune the Rashba parameter by varying temperature across the magnetic transitions which makes this material a very promising candidate for the AFM spintronics applications.    

\begin{methods}
\subsection{Sample preparation and characterization.}
PrGe single crystal was grown by the Czochralski method \cite{Das}. The sample has been characterized by XRD, magnetization, resistivity, susceptibility and specific heat measurements. 

\subsection{Angle resolved photoemission.}
 High resolution angle resolved photoemission measurements at 10~meV energy resolution and 0.2 deg angular resolution were performed at the APE beamline of Synchrotron Elettra, Italy \cite{Panaccione}. The clean surface of the PrGe single crystal was obtained by cleaving the sample in-situ in a base pressure of 9 $\times$ 10$^{-11}$ mbar.  The data have been recorded with a Scienta SES 2002 electron energy analyzer. The temperature dependent data are collected while heating using a liquid helium cooled cryostat. We have performed different data processing on the raw ARPES data (Fig.1 of supplementary information) and the data presented here are processed using Curvature method \cite{Peng}.  

\subsection{Resonant photoemission.}
 The resonant photoemission measurements was carried out at the angle-integrated PES beamline on the Indus-1 synchrotron radiation source, India \cite{Chaudhary02}. Experimental conditions are similar as reported in other references \cite{Chaudhary02, Phase2, Soma1}. The sample surface was scraped {\it in-situ} using a diamond file multiple times to obtain atomically clean surface. The BE in the photoemission spectra has been determined with reference to the Fermi level of the clean gold surface that is in electrical contact with the sample at the same experimental conditions \cite{Soma1}The intensities of the photoemission spectra were normalized to the photon flux estimated from the photo current of the post mirror at the beamline. 
 
\subsection{Density Functional Theory.}
We performed plane-wave based first-principles calculations within the density functional theory (DFT) using the generalized gradient approximation (GGA) for exchange and correlations potential using the parametrization scheme of J. P. Perdew, K. Burke and M. Enzerhof (PBE) \cite{Perdew96}. We used the Vienna ab-initio simulation package (VASP) \cite{Kresse}, which solves the Kohn-Sham equations using a plane wave expansion for the valence electron density and wave functions. The interactions between the ions and electrons are described by the  projector augmented wave (PAW) \cite{Blochl94} potentials. For our calculations, we used the PAW potential for Ge which treats ${\it 3d4s4p}$ states as valence and for Pr ${\it 5d4f6s}$ (trivalent) states were treated as valence. We performed spin-polarized calculations for both bulk and surface structures.  The expansion of electronic wave functions in plane waves was set to a kinetic energy cut-off ($E_{cutoff}$) of 350~eV for both the structures. The Brillouin-zone was sampled using Monkhorst-Pack k-point mesh \cite{Monkhorst79} of 8$\times$8$\times$8 (128 k-points in the irreducible Brillouin zone (IBZ)) and 4$\times$8$\times$1 (16 k-points in the IBZ) for the bulk and surface structure, respectively. For each structure, optimization was carried out with respect to a k-point mesh and $E_{cutoff}$ to ensure convergence of the total energy to within a precision better than 1~meV/atom. The structural relaxations were performed using the conjugate gradient algorithm until the residual forces on the atom were less than 0.01~eV/$\AA$ and stresses in the equilibrium geometry were less than 5$\times$10$^{-2}$~GPa. The total electronic energy and density of states (DOS) calculations were performed using the tetrahedron method with Bl$\ddot{o}$chl corrections \cite{Blochl94}. For calculations of spin- and site-projected DOS for Pr and Ge atoms, the Wigner-Seitz radii chosen were 2.003~$\AA$ and 1.541~$\AA$, respectively.

\subsection{Temperature dependent XRD.}
Low temperature XRD measurements were performed on the angle-dispersive x-ray diffraction beamline on the Indus-2 synchrotron radiation source. A high spectral resolution of about 1~eV at 10~KeV was achieved using Si(111) based double crystal monochromator.\cite{Sinha10} Powder XRD of PrGe single crystal was recorded at 13~KeV excitation energy by using Image plate Mar-345 detector. The photon energy and the sample to detector distance were accurately calibrated using LaB6 NIST standard. Fit2D software was used to generate the XRD pattern from the diffraction rings obtained by Image plate data. 
Low-temperature XRD measurements were also done on x-ray powder diffractometer using rotating anode type x-ray source (Cu K$_\alpha$) and CCR cryostat. 

\subsection{Magnetoresistance.}
The magnetoresistance was measured along the crystallographic [010] direction while the field was applied along the easy axis of magnetization, i.e. the [001] crystallographic direction. The measurements were performed using a Quantum Design built Physical Property Measurement System (PPMS), where the resistivity is measured by means of standard four probe method.
\end{methods}

\bibliographystyle{naturemag}


\begin{addendum}
\item {The authors wish to thank Dr. P. D. Gupta, Dr. Tapas Ganguli, Dr. G. S. Lodha and Dr. P. A. Naik of RRCAT, Indore for their constant encouragement and support. Mr. Avinash Wadikar of UGC-DAE CSR, Indore is thanked for his help in the RPES experiments. Dr. Archana Sagdeo, Mr. M. N. Singh and Mr. Anuj Upadhyay of RRCAT, Indore are thanked for the low temperature XRD measurements. Ms. Ruta Kulkarni of TIFR, Mumbai is thanked for her help in performing the MR measurements. Dr. L. S. Sharath Chandra of RRCAT, Indore is thanked for helpful discussions. Dr. J. R. Osiecki of DPCB, Sweden and Dr. K. S. R. Menon of SINP, Kolkata are thanked for providing the Igor macros for the ARPES data analysis. The authors acknowledge the support provided by DST under Indo-Italian S$\&$T POC to carry out experiments at Elettra.}

\item[Author's contributions]
S.B., A.T. and A.B. proposed the study; S.K.D. was the project coordinator; P.K.D. and A.T. prepared the single crystals; A.B., I.V., P.K.D., N.B, J.F and A.T. performed the ARPES experiments at Elettra; S.B. performed the RPES experiments at Indus-1; D.M.P. provided the experimental facility for the RPES experiments; S.B. and A.K.S. performed the temperature dependent XRD measurements at Indus-2; A.A. performed the theoretical DOS and band structure calculations; P.U.S. performed the low temperature XRD with Laboratory based source; ARPES and RPES data analysis has been carried out by S.B.; XRD and MR data analysis has been carried out by S.B and P.K.D.; S.B. prepared the manuscript with the input from all the authors.  

\item[Correspondence]
$^*$Correspondence and requests for materials should be addressed to Dr. Soma Banik~(email: soma@rrcat.gov.in, somasharath@gmail.com).

\item[Additional information]
Supplementary Information accompanies this paper. 

Present address of PKD: (1)International Centre for Theoretical Physics, Strada Costiera 11, 34100 Trieste, Italy. (2)Istituto Officina dei Materiali (IOM) - CNR, Laboratorio TASC, Area Science Park, S.S.14, Km 163.5, I-34149 Trieste, Italy.\\
Present address of SKD: Indian Institute of Technology Bombay, Powai, Mumbai 400 076, India. 
\end{addendum}

\newpage
\noindent {\large Figure Captions :}\\

\noindent Figure 1. Temperature dependent band structure of PrGe [010] surface across the magnetic transitions: (a) PM phase at 100~K showing the split bands at the $\Gamma$ point near the Fermi level ($E_F$=~0).(b) FM phase at 20~K showing the vanishing effect of the split bands. Detail band structure study at different temperature is performed to understand the nature of the split bands across the magnetic transitions and are shown in a zoomed scale around the $\Gamma$ point at (c)100~K, (d)90~K, (e)80~K, (f)70~K, (g)60~K, (h)51~K, (i)42~K, (j)33~K, (k)25~K and (l)20~K.\\

\noindent Figure 2.Comparison of the experimental VB with the theoretical DOS calculations: (a) RPES data across the Pr $4d$-$4f$ resonance are shown in a contour plot. The inset shows the CIS spectra for feature A, where the open circle is the experimental data and solid line is the fitted Fano profile. (b) the on-resonance spectra at h$\nu$=125~eV and off-resonance spectra at h$\nu$=116~eV along with the difference spectra. The yellow and red shaded regions in the difference spectra corresponds to the Pr $4f$ states and the other valence states respectively. (c) Pr $3d$ core level spectrum with the observed features $f^2$, $f^3$and $m$ marked in the spectra. In (d),(e),(f) and (g) we have shown the PDOS of Pr $4f$, Pr $5d$, Ge $4s$ and Ge $4p$ states calculated using GGA method for bulk (black filled circle) and surface slab with Pr terminated surface (blue open triangle). The crystal structure for the bulk and surface slab with the anti-parallel orientation of spins of Pr atom are also shown.(h) experimental partial density of Pr $4f$ state (as in (b), shown by green filled square) compared with the calculated partial density of Pr $4f$ states from bulk (black filled circle) and the surface slab (blue open triangle) calculations.

\noindent Figure 3.Experimental band structure compared with the stimulated energy dispersion curves and the theoretical band structure calculations: (a), (b) and (c) shows the experimental band structure at the 100~K (PM phase), 50~K (near AFM phase) and 20~K (FM phase) respectively, compared with the stimulated energy dispersion curves in the $k_y$ direction. The stimulation has been carried out with the equation described in the text. Solid and dotted lines represents hole-like and electron-like bands, respectively. E+ (red line) and E- (black line) corresponds to the bands stimulated with spin index +1 and -1, respectively. (d) and (e) shows the experimental band structure at 20~K (FM phase) and 100~K (PM phase) compared with the stimulated energy dispersion curves in the $k_x$ direction. (f) shows the crystal structure with the Pr terminated [010] surface with anti-parallel spin orientation. (g) shows the Brillouin zone in the $k_x$ and $k_y$ plane along with the high symmetry points. The red arrow shows the probable spin orientation which is along the Y-axis towards the perpendicular direction and $\delta$ is the spin-canting angle. (h) shows the theoretical band structure calculation along the $X'$-$\Gamma$-$X$ direction. (i) shows the theoretical band structure calculation along the $Z'$-$\Gamma$-$Z$ direction.

\noindent Figure 4.Characteristic of spin structures in PrGe probed by MR and XRD measurements: (a) Field dependence of the MR along the [010] direction performed at different temperatures. The MR is quantified in percentage by the following relation: MR(T,H)=~$\frac{[\rho(T,H= 0)-\rho(T,H) ]}{(\rho(T,H))}$$\times$100 $\%$. (b) Zoomed MR in the low field region to show the anisotropic MR behavior ${\it w.r.t.}$ field direction. Temperature dependent XRD patterns measured with (c) Cu K$_\alpha$ laboratory based source and (d) linearly polarized synchrotron source at 13~KeV energy. Inset in (c) shows the the rietveld refinement of the X-ray diffraction pattern measured with Cu K$_\alpha$ source at 38~K confirming the CrB type crystal structure with Cmcm space group. (e) The intensity of the peak $A$ marked in (d) is plotted as a function of temperature which shows a clear hysteresis in both PM to AFM and AFM to FM phase transitions. The hysterisis in XRD pattern is characteristic of the first order nature of the magnetic transition.

\end{document}